\documentclass[graybox]{svmult}

\usepackage{mathptmx}       
\usepackage{helvet}         
\usepackage{courier}        
\usepackage{makeidx}         
\usepackage{graphicx}        
\usepackage{multicol}        
\usepackage[bottom]{footmisc}

\usepackage{amsmath,amsfonts,amssymb}
\usepackage{graphicx}

\begin{document}

\title*{Statistical models of neural activity, criticality,\\ and Zipf's law}
\author{Martino Sorbaro, J.~Michael Herrmann and 
Matthias Hennig\\\small{University of Edinburgh, School of Informatics\\
10 Crichton St, Edinburgh, EH8 9AB, U. K.}}
\authorrunning{M.~Sorbaro, J.~M.~Herrmann, M.~H.~Hennig}
%
%
\maketitle
\date{2018}


\abstract{We discuss the connections between the observations
of critical dynamics in neuronal networks and the maximum entropy models that are
often used as statistical models of neural activity, focusing in particular
on the relation between \emph{statistical} and \emph{dynamical} criticality. We present examples of systems that are critical in one way, but not in the other, exemplifying thus the difference of the two concepts. We then discuss the emergence of Zipf laws in neural activity, verifying their presence in retinal activity under a number of different conditions. In the second part of the chapter we review connections between statistical criticality and the structure of the parameter space, 
as described by Fisher information. We note that the model-based signature of criticality, namely the divergence of specific heat, emerges independently of the dataset studied; we suggest this is compatible with previous theoretical findings.}

\section{Introduction}
The debate about criticality in neural systems began with the observation of power laws in a number of experimentally measured variables related to neural activity. The first experimental observation of neuronal avalanches~\cite{beggs2003neuronal} found that their size distribution follows a power law with exponent of about $-3/2$, and their duration distribution follows one of exponent near $-2$  in cortical slices. These values are compatible with the exponents expected in critical branching processes  --- a well-studied topic in the field of complex systems physics~\cite{Athreyab}. 
Similar observations have been consistently reported in literature; moreover, the presence of power-law avalanche statistics was found to be theoretically justified by functional arguments on numerous occasions \cite{beggs2008criticality, shew2009neuronal, shew2013functional, gautam2015maximizing, vazquez2017stochastic}, and was shown to differ in different brain states~\cite{priesemann2013neuronal, hahn2017spontaneous}. 
For an excellent high-level discussion of the topic, see~\cite{beggs2012being}.

An equally interesting instance of a power law is the finding that, in the population statistics of a neural network's activity, 
the rank of a state (the first being the most frequently observed, and so on) and its frequency are inversely proportional. This  phenomenon, known as Zipf's law,
was first observed by Auerbach in 1913: 
``If one sorts individuals by a given property
in a descending fashion and stops doing so at 
rank $n_1$, or at $n_2$, or generally at rank $n_x$,
where the property has gone down to values $p_1$, $p_2$,
$p_x$, then a certain law exists between $n_x$ and $p_x$.
In our case, this law is especially simple, it is expressed by
the formula: 
$n_x \cdot p_x = \rm{constant}$''~\cite{Auerbach1913}.
This author already alluded to the possibility of more complex forms of the same law (e.g.~for the distribution of wealth), but did not 
speculate why it assumes its simplest form in the studied data.
In the mid-1930s, the American linguist George Kingsley Zipf 
discovered that the frequency of occurrence of words in
Joyce's Ulysses and American newspapers
follows the same law~\cite{Zipf1949}, which is today
called Zipf's law: the frequency of each word decays 
as a power law of its frequency rank. 
After Auerbach's original example, 
city sizes~\cite{gabaix1999zipf,jiang2011zipf}, Zipf's law was confirmed in a variety of fields, 
including citation 
counts in scientific literature~\cite{redner1998popular}, 
earthquake magnitudes, wealth, solar flare size, number of emails and phone calls received, and many others \cite{newman2005power} . 

Over the years several attempt have been made to understand Zipf's law.
Zipf himself explains it by the \emph{principle of 
least effort}: If words are stored in a linear array, then the low-frequency
items are optimally located in a more distant place than more often used ones.
The product of distance and frequency can be considered as a measure 
of the effort necessary to retrieve the word which he claims to be a constant. However, the assumption
of an array, where the effort needed for retrieving an item is linear, which is necessary in order to obtain an inverse relationship rather than a general power law, seems unnatural when considering how items are stored in a neural network. 

Another potential cause for the law can be seen in the idea of
preferential attachment~\cite{Barabasi1999}. If, for example, the probability 
to move to or away from a city is assumed to be independent of its size, then 
Zipf's law for city sizes emerges. Other assumptions have been 
discussed and been shown to provide a better match for the 
distribution of city sizes~\cite{Vitanov2015}, but again this 
may not easily carry over to states in a neural network.

Li~\cite{Li1992} demonstrated that the words of 
an artificial language that simply consists of randomly chosen letters 
including a space sign tend to obey Zipf's law.
However, the `space' sign, which separates words in Li's approach, plays
no such role in the analysis of neural data.

The authors of Ref.~\cite{aitchison2016zipf} aim at an explanation
of Zipf's law by the existence of latent variables. Differently from the 
above attempts, this study is directly relevant for the 
analysis of neural activity. It also subsumes the scheme 
proposed by Li~\cite{Li1992}.

Although all of these attempts have their interest, there is some agreement
that a deeper understanding is still lacking.
In addition, there seems to be no clear 
justification on why criticality in the \emph{statistical} sense and Zipf's 
law have been observed in neural data, or what brain function might 
benefit from it. It is interesting in this context that 
Zipf's law is a system property, i.e.~it depends on the number of elements
in the system and does not automatically 
apply to subset or unions of Zipfian sets. It can not be reduced 
to the mere presence of a particular probability distribution (such as $P(x)\sim x^{-2}$), but requires a conditional sampling procedure to be reproduced in
a simulation~\cite{Cristelli2012}. 
The observations in neural data as well as a number of unsolved 
problems with this subject
make it a very interesting subject of further investigation.

In what follows, we will discuss the connections between the observations
of critical dynamics and maximum entropy models that are
often used as statistical models of neural activity, 
reviewing the recent literature on the matter, and debate the possible relationship between this \emph{statistical} concept of criticality and the \emph{dynamical} criticality related to avalanche statistics. First, we will illustrate the concept of Zipf's distribution, its origin, and its applicability to neural data.
 In section \ref{sec:statmodels}, we will introduce maximum entropy models, and show their connection to criticality and Zipf's law. 
In section \ref{sec:statcrit}, we will make three observations
that emphasise the difference between statistical and dynamical criticality:
(\ref{subsec:Eurich}) a system that shows dynamical, but not statistical criticality, 
(\ref{subsec:Fitting}) the process of fitting an energy-based model, and 
(\ref{subsec:Ret}) the application of the theory of a large-scale corpus of biological data, where Zipf's law appears to hold, although the system is not
dynamically critical.
In section \ref{sec:fim}, we will show connections between statistical criticality and the structure of the parameter space, as described by Fisher information. Finally, in Section~\ref{Discussion}, we will return to the question 
debate whether there is a relationship between statistical and dynamical criticality and conclude with an outlook on the problem.

\section{Statistical description of spike trains}
\label{sec:statmodels}
\subsection{Zipf's law in neural data}

\begin{figure}[t]
\begin{center}
	\includegraphics[width=0.5\textwidth]{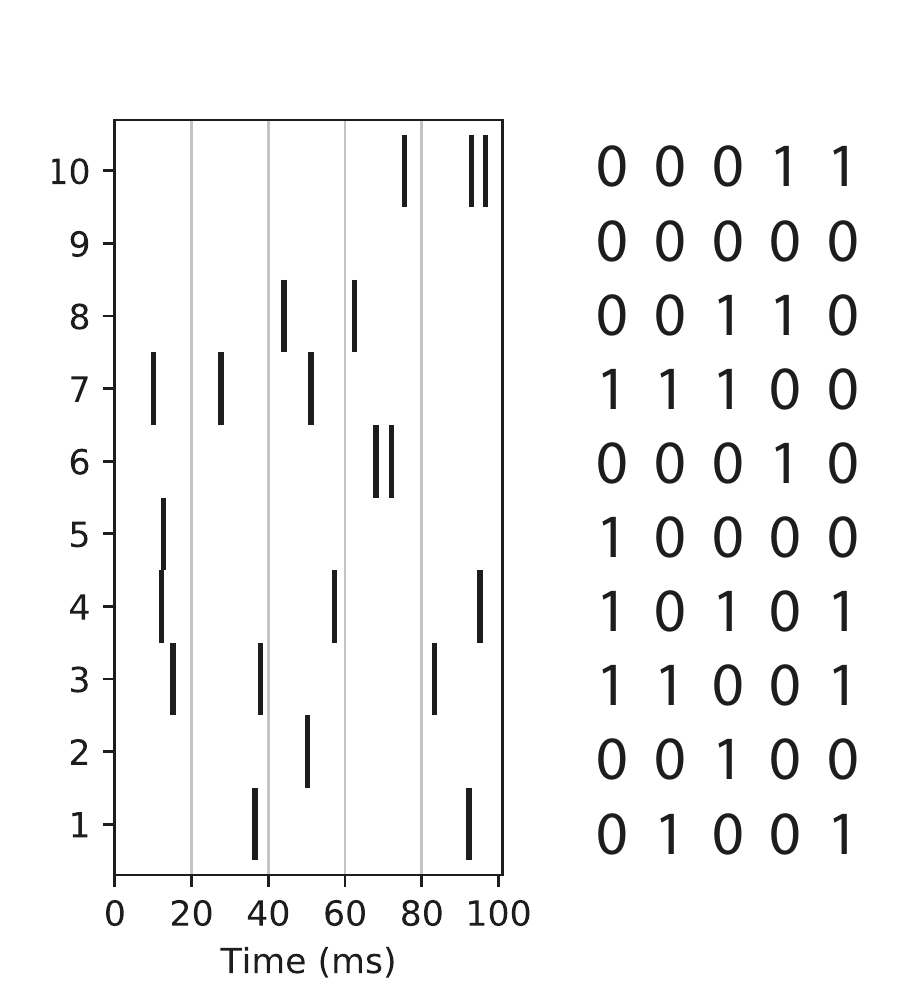}
	\caption{Digitisation of spike trains from 10 neurons into a boolean matrix with bin size $\delta t = 20$ ms.}
	\label{fig:spktrain}
\end{center}
\end{figure}

For the specific case of neural activity, Zipf's law refers to the rank-probability law for the occurrence of each possible \emph{pattern} of activity, which has been observed to follow a power law in the same sense as for words in the English language \cite{tkacik2015thermodynamics}. 
To understand what we mean by \emph{pattern} or \emph{state}, we need to adopt a simplified way of representing spike trains that we can call \emph{digital}: discretising time in bins of equal size $\delta t$, we can define a Boolean variable
\begin{equation}
\sigma_n(t) = \begin{cases} 1 & \text{ if neuron $n$ spikes between $t$ and $t+\delta t$ }\\
					     0 & \text{ otherwise.} \end{cases}\label{Boolean} \end{equation}
 At any given time, then, the population activity is described by a \emph{codeword}
 \[ \boldsymbol{\sigma}(t) = (\sigma_1(t),\dots,\sigma_N(t)) \]
which describes, up to a precision of $\delta t$, the spiking state of the $N$ neurons considered (Figure \ref{fig:spktrain}). Modelling the system statistically, in this framework, means giving a full account of the probability of each possible codeword to appear. Note that, typically, we are not concerned with the dynamics of the system, and we disregard temporal correlations on scales larger than $\delta t$: this approach is suited to describe short-time correlations across space 
or properties of the encoding. Needless to say, the choice of $\delta t$ can have important consequences on the results: in the limit of very large bin size, the pattern where all neurons fire simultaneously will be the only one to be observed; in the opposite limit of small $\delta t$, the \emph{silent} pattern will be the most common, patterns with a single active neuron arbitrarily rare, and multi-neuron patterns absent. The results we discuss hold for bin sizes of the order of 5--20 ms, i.e. of the same order of magnitude as the typical correlation length between neurons; this value is commonly adopted in the literature \cite{schneidman2006weak}.

To understand why Zipf laws are considered a signature of criticality, we will now illustrate the relationship, exposed by recent literature, between them and the critical points of models that have been used to describe neural activity, and are well known in physics.

\subsection{Statistical modelling}
The activity patterns of individual neurons and neural networks invariably display stochastic characteristics. A common approach, which we can call \emph{top-down}, of modelling the nature of this activity is to make (simplifying) assumptions on the actual workings of neurons, synapses, and networks, in order to set up a computational model the results of which can then be compared with experimental observations. 
A large part, perhaps the largest, of computational neuroscience is based on this paradigm, predominantly by simulations of spiking neural networks. 

Here, on the contrary, we are concerned with what we call \emph{bottom-up} modelling, which seeks to infer properties of neural activity in an entirely data-driven way. Understanding the correlation structure, the distribution of firing rates, or the repetition of identical patterns from experimental data are examples of this approach. In other words, the data is described in terms of probabilities and other statistical descriptors, instead of parameters directly implied by the biological or physical theory.

In the bottom-up approach, a very broad family of models is available. We will restrict ourselves to \emph{energy-based statistical models}, a number of models developed in the last decade which adopt a log-linear relation between probability and state variables.
In an \emph{energy-based} model probabilities are expressed 
in terms of an energy function $E$, in analogy with statistical physics:
$$ P(\boldsymbol\sigma) = \frac{1}{Z} e^{-E(\boldsymbol\sigma)},$$
where $Z$ is the relevant normalisation factor.
Many energy-based models used in neuroscience adopt the aforementioned \emph{digital} description of spike trains in terms of binary variables: we will focus on these. In this case, for $N$ neurons, $\boldsymbol\sigma$ can take $2^N$ different values, and determining the full population probability distribution requires specifying $2^N$ probabilities, which is an unrealistic task even for modest population sizes. Assumptions on the analytical form of the distribution are therefore required in order to infer a complete distribution from a relatively small number of samples.

\subsection{Maximum entropy models}
The first, and perhaps more elegant, strategy developed to this end is to adopt a \emph{maximum entropy} approach~\cite{jaynes1957information}, in which one first selects what features of the data should be exactly reproduced, and determines then the highest-entropy probability distribution consistent with those constraints. Schneidman et al.~\cite{schneidman2006weak} and Shlens et al.~\cite{shlens2006structure} first applied this approach to neural data, using a Pairwise Maximum Entropy (PME) model, which exactly fits all $\langle \sigma_i \rangle$ and $\langle \sigma_i \sigma_j\rangle$, i.e.\ firing rates and pairwise correlations. Indeed, the question behind that research was primarily related to the importance of correlations in the vertebrate retina, including the study of higher-order interactions.

The PME probability distribution over all codewords has the following form:
\begin{equation} \label{eqn:ising} P(\boldsymbol\sigma) = \frac{1}{Z(h, J)} \exp\left( {\sum_{i=1}^N h_i\sigma_i + \sum_{i \neq j}J_{ij}\sigma_i\sigma_j} \right).\end{equation}
The expression above is mathematically identical, in statistical physics, to that of the canonical ensemble for the Ising model with arbitrary couplings, a generalisation of the model originally used to describe ferromagnetism in solids~\cite{ising1925beitrag}.

By definition, a successful fit of a PME model correctly reproduces all firing rates and pairwise correlations present in the data from the considered 
neural population. Fitting based solely on second-order statistics does not imply that third-order correlations and other statistical measures are correctly reproduced. 
Reports that higher-order correlations are largely irrelevant were thus very surprising~\cite{schneidman2006weak,shlens2006structure,Tang2008}, although these observation may be restricted to low activity and high pairwise correlations~\cite{Yu2011}.
Assessing whether a maximum entropy model can capture additional statistics of the data provides a source of interpretability: If, say, a PME model can account for third-order correlations, then the latter are not constrained further by the data. If, conversely, third-order correlations diverge from the PME prediction, we learn that the neural activity uses higher-order statistics to encode information. Whether this is the case depends on the system and on the distance between the neurons considered \cite{ohiorhenuan2010sparse}.

Several attempts have been made at improving the quality of the fit of statistical models, using different features as known statistics.
The generalisation of equation (\ref{eqn:ising})
\[ P(\boldsymbol\sigma) = \frac{1}{Z(h, J)} \exp\left( {\sum_{i=1}^N h_i\sigma_i + \sum_{i, j}J^{(2)}_{ij}\sigma_i\sigma_j} +
\sum_{i, j, k}J^{(3)}_{ijk}\sigma_i\sigma_j\sigma_k  + \dots \right)  \]
can describe any probability distribution of binary variables exactly. However, finding the values of $J^{(n)}$ is computationally expensive and the benefits typically do not outweigh the costs for $n \geq 3$.

As a different way to assess at least some aspects of the higher-order statistics, we can consider, for instance, the probability distribution of the number of neurons firing in a time bin, $p(K)$, where $K(t) = \sum_{i=1}^N \sigma_i$, was used as a target. This can be introduced as a further constraint in a maximum entropy model in combination with firing rates and pairwise correlations, leading to the $K$-\emph{pairwise} model~\cite{tkacik2014searching,mora2015dynamical}. 
It typically produces significantly better fits than a pure PME, and is much less computationally expensive than attempting to fit higher order cumulants. 
Another related approach, the \emph{population tracking model}, fits $p(K)$ together with the conditional probabilities $P(\sigma_i=1| K)$ of each neuron firing, given the current population firing rate, providing a lightweight and interpretable model~\cite{odonnell2016population}.

An example of an energy-based model which does not rely on the maximum entropy principle, finally, is to use a restricted or semi-restricted Boltzmann machine (RBM/sRBM). Despite not directly aiming at fitting correlations, $p(K)$, and cumulants, as a maximum-entropy model would, RBMs were shown to perform at least comparably well in fitting all these aspects~\cite{koster2014modeling}. An advantage is that their complexity can be tuned, offering a choice of various degrees of accuracy and the corresponding computational costs.
Additionally, \emph{contrastive divergence}, the algorithm used for fitting, is an approximate but relatively fast and reliable algorithm, which lets one fit the simultaneous activity of a large number of units (up to several hundreds, whereas the exact learning algorithm for a PME model is not usable in practice over $N\approx 40$, although more efficient methods have been studied). Finally, RBMs can be interpretable models, specifically by studying the roles taken by hidden units.

Although a detailed discussion is beyond the scope of this chapter, we should at least mention the efforts to reproduce the time dynamics of the system, so that the statistical model fits both the distribution of single-time bin patterns and the conditional distribution of the pattern given the pattern in the previous time bin~\cite{marre2009prediction, nasser2013spatio, gardella2018blindfold}. 

\subsection{Phase transitions in models}

Although this may initially seem less relevant from the point of view of  
research in neuroscience, we should remind ourselves that the Ising model is 
one of the earliest and most commonly studied paradigms of a phase transition. 
To understand its behaviour, let us make the temperature dependence of equation (\ref{eqn:ising}) explicit:
\begin{equation} P_T(\boldsymbol\sigma) = \frac{1}{Z(h/T, J/T)} \exp\left( \frac{1}{T}\left[{\sum_{i=1}^N h_i\sigma_i + \sum_{i \neq j}J_{ij}\sigma_i\sigma_j} \right]\right)\label{Ising_T}\end{equation} Note that, in the high temperature limit, this converges to a uniform probability:
\[ P_{T\to \infty}(\boldsymbol\sigma) = \frac{1}{2^N}, \quad \forall\, \boldsymbol\sigma. \]
Conversely, when $T\to0$, only a small number of states with non-zero probability survive, the others becoming infinitely rare. If a system that obeys 
$P_{T\to 0}(\boldsymbol\sigma)$ is perturbed in any way, it will
eventually converge to this stable set, under any reasonable dynamics. In other words, the distribution becomes, in the zero temperature limit, a finite set of stable attractors, the same as the stationary distribution of a Hopfield network~\cite{Hopfield1982}, where the attractors play the role of memory patterns.

Clearly, neither of the two limiting cases can be a realistic description of neural statistics, and the truth lays in between them, in a regime where the model is much more informative. The physics literature shows that there is sharp phase transition between a \emph{disordered} phase and a \emph{spin glass} phase, with the exact location of the critical point depending on the statistics of $h$ and $J$ \cite{nishimori2001statistical}. It is then a natural question to ask whether the Ising model that results from a fit to neural activity is in one of the two phases, or poised near the critical point, and whether this relates to other concepts of criticality in neural systems.

The divergence of specific heat, also called heat capacity, in a macroscopic system is a classic signature of discontinuity in the properties of the system upon variation of a single parameter, typically temperature (generalisations of this idea will be discussed in the next section). The most classic example is the case of a change in the state of matter --- solid to liquid, liquid to gas, etc.~--- where an infinitesimal change in temperature through the critical point requires a finite amount of energy (the \emph{latent heat}). This is equally true for spin systems of the form we examined above. Tka\v cik et al.~\cite{tkacik2015thermodynamics} fitted a model of the form (\ref{eqn:ising}) to binned spike trains from recordings of the salamander retina subjected to movies of naturalistic stimuli. They varied the temperature of the model around $T=1$ (this value corresponding to the fit to neural data), and studied the specific heat as a function of $T$ for an increasing number of neurons.

Their result clearly showed a peak in the specific heat of their models, with the peak temperature approaching $T{=}1$ as $N$ is increased. This is evidence that $T{=}1$ coincides with the critical point, and therefore, the model is poised at criticality for parameter values exactly corresponding to those that fit the neural data. Similar observations were independently repeated, e.g. in~\cite{mora2015dynamical, nonnenmacher2017signatures, hahn2017spontaneous}, generating a debate on the nature of this observation and its biological interpretation, as will be discussed in later sections.

\subsection{Model criticality and Zipf's law}
Zipf laws can be related to statistical criticality in the sense of models, as shown in~\cite{tkacik2015thermodynamics} (supplementary information), as follows. Call $p_1,...,p_k,...,p_{2^N}$ the probability of occurrence for each of the $2^N$ possible codewords. In statistical physics, microcanonical entropy can be defined as $S = \log\Omega$, where $\Omega = \Omega(E)$ is the number of states with energy lower than $E$. On the other hand, the energy level associated with a pattern is a function of its probability:
\[ E_k = -\log p_k + \text{const.} \]

Now, Zipf's law states that, for every pattern, its rank $r_k \propto 1/p_k$. In the notation used above, note that $r_k = \Omega(E_k)$. Therefore, Zipf's law implies
\begin{eqnarray}
\log p_k &=& -\log r_k + \text{const.} \cr
 E_k &=& S_k + \text{const.} 
\end{eqnarray}
If the above linear relation holds, then $d^2S/dE^2{=}0$. Since both specific heat and the variance of energy are inversely proportional to $d^2S/dE^2$, these thermodynamic quantities diverge.
This is the classic signature of a second order phase transition.

The rank-probability relation defined by Zipf's law, therefore, is a model-in\-de\-pendent way of showing criticality in this statistical sense. Its appearance guarantees the divergence of the specific heat of a PME model fit to the same data, but does not require complex and computationally expensive fitting procedures, and relies only on the statistical properties of the data.

\section{Statistical and dynamical criticality}
\label{sec:statcrit}
As we have mentioned, most energy-based models do not account for dynamics, as they are concerned only with fitting a single-time bin distribution. The formulation of the Ising model in physics describes a stationary distribution and does not include any dynamics. Transition probabilities from a state to another can be added through additional assumptions about the dynamics of the system. For instance, Glauber dynamics~\cite{Glauber1963} generates a Markov chain whose stationary distribution coincides with the distribution (\ref{eqn:ising}). This is useful, for example, when sampling states from that probability distribution. Avalanches can be observed in high-dimensional Ising systems when they are driven out of stationarity by a change in temperature or applied magnetic field. Therefore, it is not expected that maximum entropy models reproduce any aspect related to avalanche dynamics.

It is not clear a priori, then, whether the observation of Zipf laws and diverging specific heats should be related to power-laws in the dynamics. In fact, finding a connection between the two concepts seems challenging. In the next sections, we will provide examples of how the two might be entirely distinct, which prompts questions on the nature, meaning, and relevance of statistical criticality.

\subsection{The Eurich model is dynamically, but not statistically critical\label{subsec:Eurich}}
\begin{figure}[t]
\begin{center}
	\includegraphics[width=\textwidth]{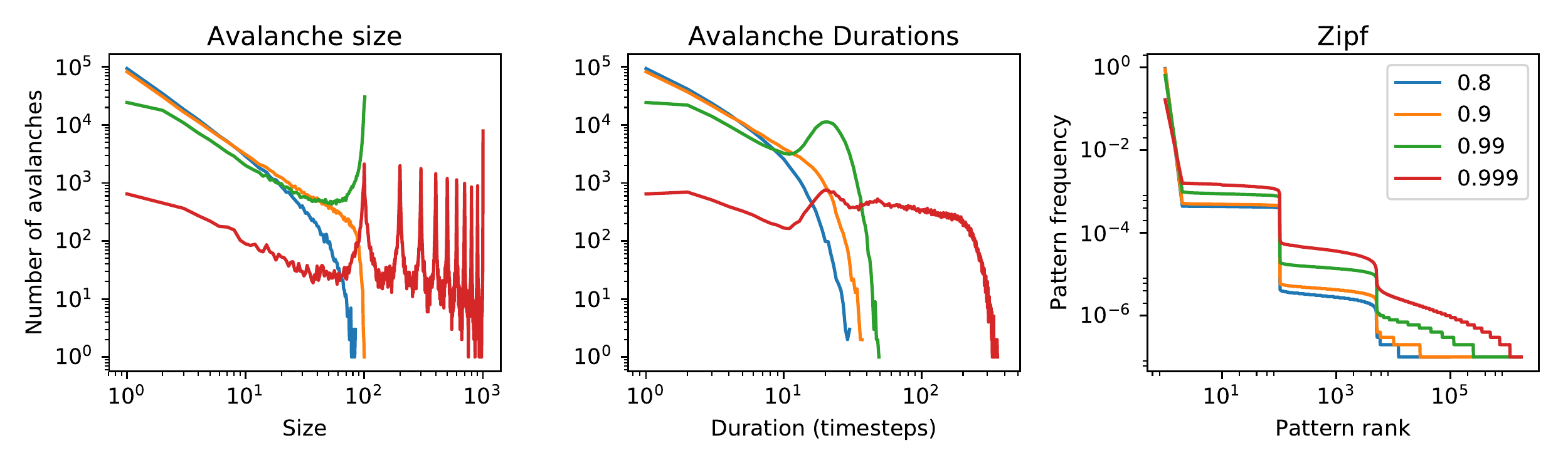}
	\caption{Avalanche statistics compared to Zipf law for the Eurich model in different regimes~\cite{eurich2002finite}. Here $N=100$, and the critical point is at $\alpha_c=0.90$: in this case, the avalanche size and duration distributions most closely approach a power law with exponent $-3/2$, except for a cutoff due to the finite size. The subcritical case (blue), on the other hand, shows short-tailed distributions, and the supercritical cases (green and red) exhibit respectively one or many peaks at large values. In all cases, the Zipf plot does not show power-law dependence. Note that the smoothing of the step-like function is due to the finite sample size.}
	\label{fig:eurich}
\end{center}
\end{figure}

For a discussion of the relationship between the two concepts of criticality, it is interesting to consider the Eurich model for neural avalanches as a ``testbed''~\cite{eurich2002finite}. It is mathematically well-understood and can be 
conveniently tuned, because the parameter values for 
\hbox{(quasi-)}critical as well as sub- or super-critical 
behaviour are known analytically (even for finite systems).
The very definition of the model is such that all neurons have identical properties, and the same for pairs, triplets, etc., of neurons. As a consequence, all $N$ patterns with exactly one active neuron appear with the same frequency; all $N(N-1)/2$ patterns with exactly two active neurons appear with the same frequency, and so on, giving the rank-probability plot a step-like appearance, which cannot follow a Zipf law (Figure \ref{fig:eurich}).

It is tempting to consider the tail of the rank distribution.
Although the number of states increases with the activity (for 
neurally plausible activity levels), their probability decreases
strongly if the firing rate (per time bin) is low. Therefore,
steps will disappear for higher ranks, which may or may not 
produce a power-law-like behaviour. 
However, in the statistical approach, typically small values of
$N\delta t$  (see equation \ref{Boolean}) are used, such that 
the potentially Zipf-like tail 
(figure~\ref{fig:eurich}, right) will be statistically 
irrelevant.  It has also been claimed that 
the statistical approach is most likely 
restricted to low-activity patterns~\cite{Yu2011}.

Note that the rank curve would be less step-like if some form of heterogeneity is introduced, as opposed to the complete symmetry between neurons that characterises the Eurich model. However, this would amount to an additional assumption that is, as the Eurich model shows, not necessary for criticality. In particular, we would need to assume that the patterns with a single active neuron already follow Zipf's law. This is a particular, but not unreasonable assumption, as this can be expected, for example, in a scale-free neural network.
 In fact, we cannot rule out that a Zipf profile, or at least an approximation, could be found just by tuning the distribution of firing rates, even in the absence of correlations. Such a finding would entirely rule out any relation to dynamical criticality, which appears exclusively as a consequence of emergent phenomena deriving from complex interactions. However, it would still require a specific distribution of firing rates among the neurons, an assumption that in itself would prompt questions about its functional reasons. Firing rate distributions have been studied extensively \cite{mizuseki2013preconfigured}, and found to be highly skewed, with a small fraction of neurons responsible for the majority of emitted spikes. A clear theory on why this is an advantage for the encoding is still missing. In section \ref{sec:retina}, we will consider a case in which the Zipf relation holds even when correlations are destroyed, which suggests the long-tailed firing rate distribution is sufficient for it to hold.

\subsection{Fitting energy-based models to critical activity\label{subsec:Fitting}}
\label{sec:rbm}

A natural way of checking if dynamical and statistical criticality are related could involve fitting a statistical model to neural models that exhibit various kinds of dynamics, and can be tuned to a supercritical (noisy), subcritical or critical regime. This was one of the goals of in Ref.~\cite{hahn2017spontaneous}. 
The authors identified five different dynamical states of the cat and monkey cortex, studied their avalanche statistics, and evaluated the temperatures corresponding to peak specific heats of Ising models fitted to each dataset. The results did show a small but significant relationship between avalanche dynamics and specific heat peak location.
However, it should be noted that some of the results in this work were obtained with small datasets of six neurons only, which may not offer insight on what happens in the thermodynamic limit.


We attempted a similar task fitting Bernoulli RBMs to the activity generated by a tunable model of a neural network, similar to the binary, non-leaky, integrate and fire model used by~\cite{gautam2015maximizing}. In this model, the strength of inhibition can be tuned, leading to a network with low, random-looking activity and a short-tailed avalanche distribution (high inhibition); or a network generating activity in large bursts (low or no inhibition). The critical regime lies in between the two. Details about the implementation of the model are given below.

We found that although the absolute value of the peak does depend on the correlations of the data, its location is always at a temperature near $T=1$ (which is the value corresponding to the original fit), and further approaches this temperature as the number of units increases, i.e. in the thermodynamic limit. These results are compatible with what was shown by \cite{nonnenmacher2017signatures}, using a different dataset, for the $K$-pairwise model.

It seems, then, not only that the statistical model that we fitted does not accurately detect criticality in the dynamical sense, but it also exhibits statistical criticality no matter the dataset it was fitted to. This implies, on the one hand, that the dynamical criticality of a dataset and the statistical criticality of a model fitted to it are unrelated, and, on the other hand, that a model fitted to datasets of very different nature all tend to exhibit statistical criticality. This is compatible with an argument that was put forward by theoreticians \cite{mastromatteo2011criticality}, as will be discussed in section \ref{sec:fishcrit}.

\begin{figure}[t]
\begin{center}
	\includegraphics[width=\textwidth]{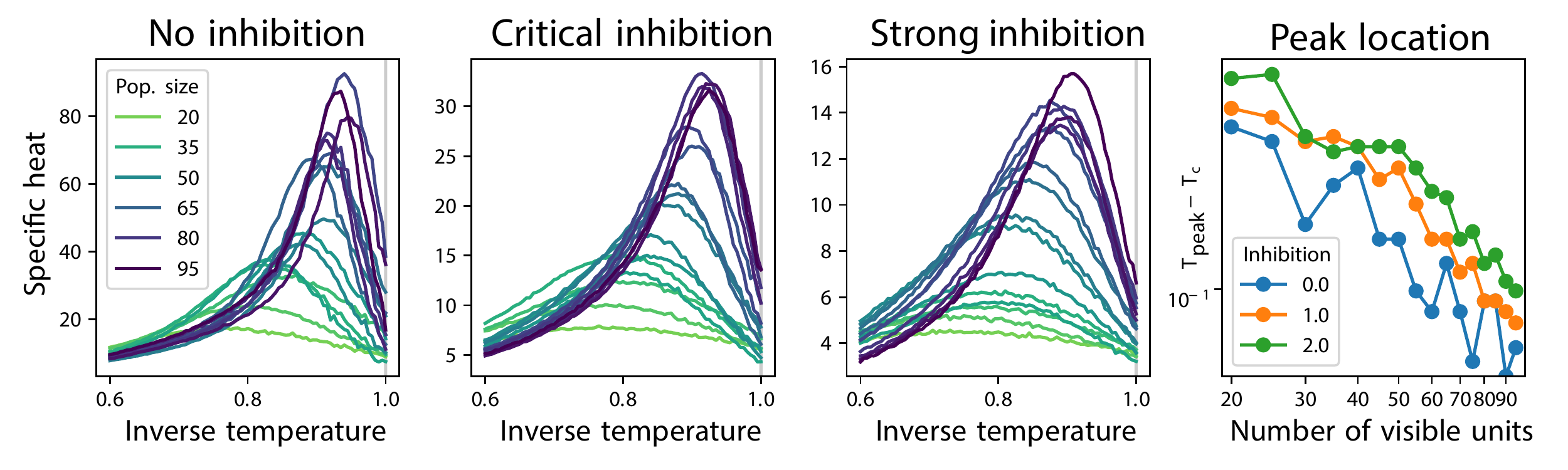}
	\caption{RBM specific heats peak when fitted to a variety of datasets, with the peak approaching the temperature of the fit as the model size increases. RBMs were fitted to simulated data with different avalanche statistics: supercritical (left), critical (centre) and subcritical (right).}
	\label{fig:binneur_cvs}
\end{center}
\end{figure}

\paragraph*{Methods: network model}
The binary neuron model used for the simulations is similar to the one presented by \cite{gautam2015maximizing}. In this model, each neuron has a probability of firing given by a weighted sum of its inputs, divided by a factor dependent on its own firing history. A fifth of all neurons are inhibitory, while the rest are excitatory; the value of inhibition was tuned to 0.0, 1.0, or 2.0, to enforce different regimes, corresponding to different correlations and avalanche statistics. While in the original work the connectivity was all-to-all, with weights drawn from a uniform distribution, we modified it to identical couplings, but set on a network with scale-free degree distribution. This enforced larger variability of firing rates between neurons; both experimental evidence and the theory in \cite{larremore2011predicting} suggest this choice does not affect the location of the critical point. We simulated 1000 neurons, from which we took subsets of the sizes required for analysis, for 1 million time steps (conventionally taken to equal 1 ms). The resulting activity was re-binned in 5 ms bins, to reduce sparseness.

\paragraph*{Methods: RBM specific heats}
As we will argue in section \ref{sec:fishcrit}, the direction that best indicates the critical point coincides with the fist eigenvector (the one corresponding to the largest eigenvalue) of the Fisher information tensor. However, in practice, this is never orthogonal to the direction of increasing/decreasing temperature: thus, varying temperature is an acceptable way to look for a phase transition.

In statistical physics, the general expression for the probability of a pattern in an energy-based model at temperature $T$ is
\[ P_T(x) = \frac{1}{Z(T)}e^{-E(x)/T}. \]
The expression for the energy in the case of RBMs is 
\[ E(\mathbf{v}, \mathbf{h}) = - \mathbf{a}^\intercal\mathbf{v} - \mathbf{b}^\intercal\mathbf{h} - \mathbf{v}^\intercal J\mathbf{h}. \]
Where $\mathbf{v}$ and $\mathbf{h}$ are vectors of visible and hidden binary variables respectively. Since this expression is linear in the parameters $a_i, b_j$ and $J_{ij}$ for all $i, j$, changing the temperature of a model coincides with rescaling these parameters by a linear factor $\beta = 1/T$. In the following, we have adopted the standard strategy of fitting an RBM to neural data, obtaining values for its parameters, and then rescaling them --- this means $T=1$ (no rescaling) coincides with the parameters as they were fitted, the values corresponding to a model that correctly reproduces the given data. Fits were obtained by 1-step persistent contrastive divergence.

We can then compute the specific heats at different temperatures. The marginal probability of $\mathbf{v}$ is
\[ P_T(\mathbf{v}) = 
\frac{1}{Z}\sum_{h_{1...N} = 0, 1}e^{\frac{1}{T}(\mathbf{a}^\intercal\mathbf{v} + \mathbf{b}^\intercal\mathbf{h}+\mathbf{v}^\intercal J\mathbf{h})} 
= \frac{ e^{\mathbf{a}^\intercal\mathbf{v}/T}}{Z}  \prod_{j=1}^{N_h}\left(1 + e^{b_j/T + (\mathbf{v}^\intercal J)_j/T}\right) \]
Disregarding an additive constant, the energy of a visible pattern can be expressed as the logarithm:
\[ E_T(\mathbf{v}) = \frac{\mathbf{a}^\intercal\mathbf{v}}{T} + \sum_{j=1}^{N_h}\log\left( 1 + e^{b_j/T+ (\mathbf{v}^\intercal J)_j/T} \right)  \]
In accordance with statistical physics, we can define
\[ c(T) = \frac{\operatorname{Var}(E_T)}{N_vT^2}. \]
This quantity can be computed from a sample.  For each temperature value, in each dataset, we extracted 20 chains of 2000 samples, taken every 10 steps in order to reduce spurious correlations, and computed $c(T)$ by the expression above.




\subsection{Retinal activity obeys Zipf's law\label{subsec:Ret}, but is not dynamically critical}
\label{sec:retina}
The mammalian retina, a system that is often chosen when studying the statistics of neural activity, and whose encoding and dynamical properties are well known, is an example of the opposite case: It was the first system in which statistical criticality was observed, but it does not exhibit dynamical criticality.

\begin{figure}[t]
\begin{center}
	\includegraphics[width=\textwidth]{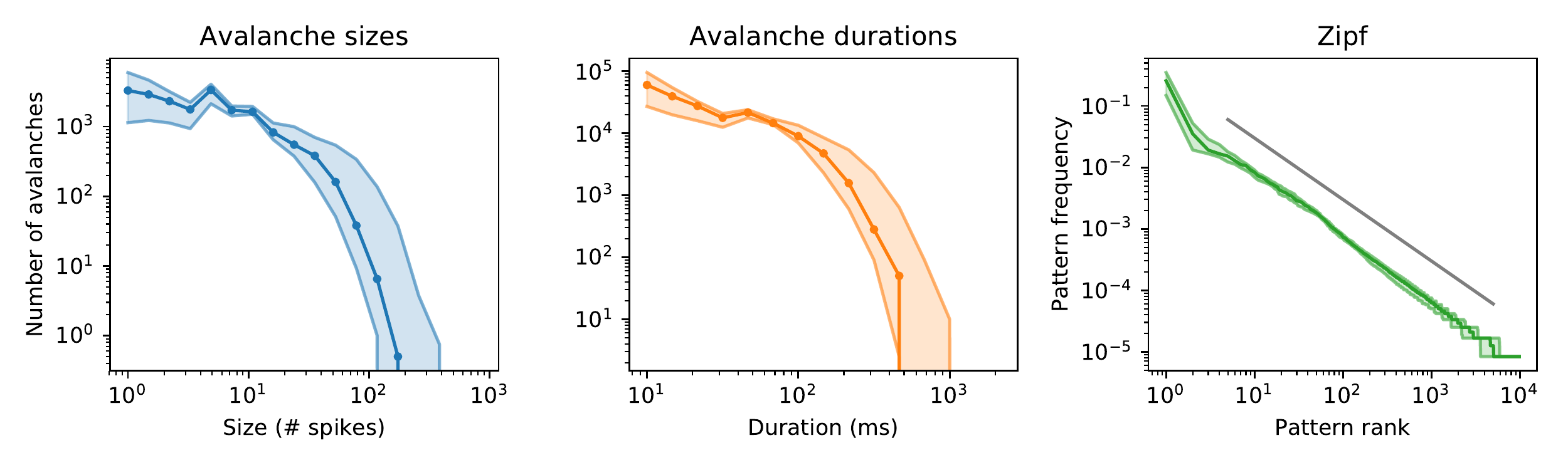}
	\caption{Avalanche statistics compared to Zipf law in the neural activity of a healthy, adult (postnatal day 91) mouse retina stimulated by projection of a white noise checkerboard pattern. The detection of avalanches and the pattern count were repeated over 30 sets of 100 neighbouring neurons, each of which was recorded for 20 minutes. The sets may overlap. The solid lines are medians over these sets; the shaded area is delimited by the first and third quartiles. The grey line in the rightmost plot is for comparison with Zipf's law. The data were made available by G. Hilgen and E. Sernagor, University of Newcastle. We refer to \cite{hilgen2017unsupervised} for experimental and data analysis methods.}
	\label{fig:zipf_retina}
\end{center}
\end{figure}

Avalanches arise in the mammalian retina only during the period of development: for mice, in the first few days after birth, before eye opening, when the retina does not respond to light and the network activates spontaneously. During this stage, the activity of the retina consists of the so-called \emph{retinal waves}, which are effectively power-law distributed avalanches. Direct comparison with a computational model showed that these are indeed the signature of a critical state between locally and globally connected activity~\cite{hennig2009early}. However, these disappear in a functional retina: 
Figure \ref{fig:zipf_retina} shows the statistics of a 20-minute recording of an untreated, adult mouse retina under an uncorrelated black-and-white checkerboard stimulation. It is evident that the avalanche statistics is short-tailed, and, at the same time, the probability-rank plot of pattern frequencies is well compatible with a Zipf law. Note that correlations between the activities of retinal ganglion cells change significantly with the statistics of the stimulus, and the avalanche statistics will consequently appear different.
The example of adult retinas is complementary to Section~\ref{subsec:Eurich} in the sense that, here, a system that does not show dynamical criticality can well obey Zipf's law.

It is worth mentioning that the observation of Zipf law in retinas is very robust to a number of external factors. We found no significant differences in the rank-frequency plots of patterns observed when the retina was treated with bicuculline (a GABA blocker) compared to a control; analogously for retinas under stimuli characterised by very different level of spatial correlations.

 \begin{figure}[t]
	\begin{center}
		\includegraphics[width=0.24\textwidth]{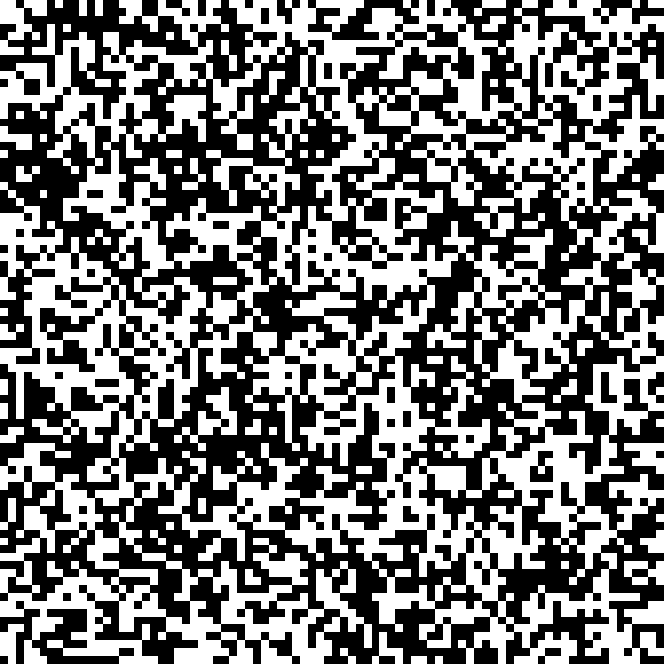}
		\includegraphics[width=0.24\textwidth]{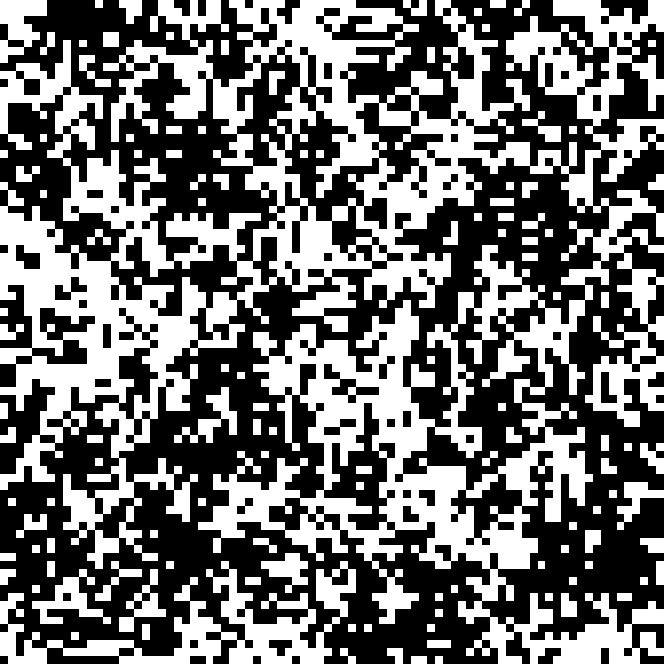}
		\includegraphics[width=0.24\textwidth]{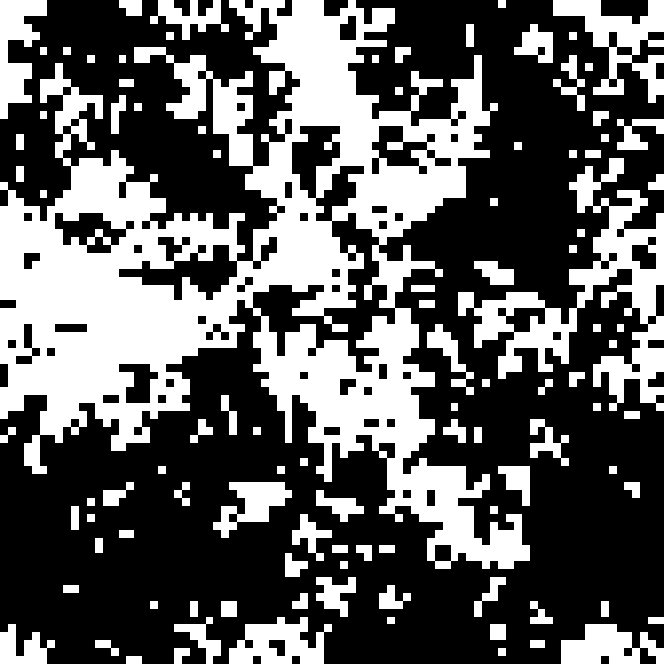}
		\includegraphics[width=0.24\textwidth]{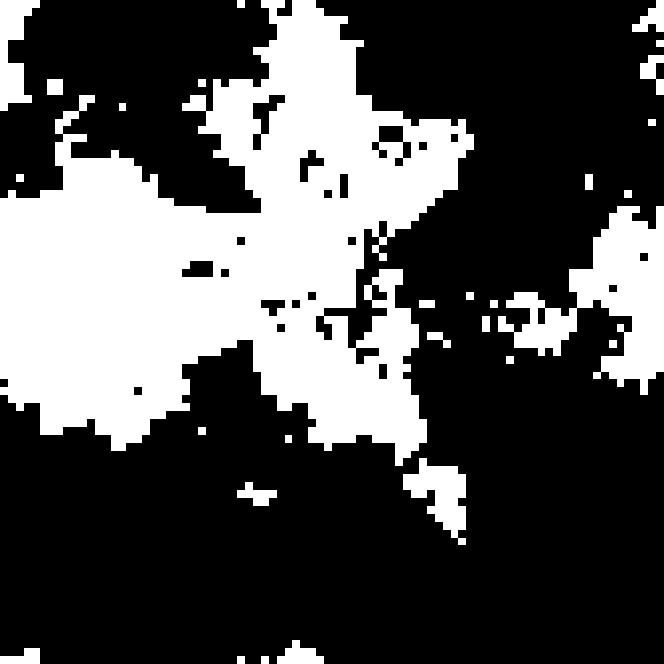}
		\caption{Examples of stimulation frames. Correlations increase from left to right (dataset ID 0 to 3): the frequency spectra follow $f^{-a}$ with $a= 0.5, 1.0, 1.5, 2.0$, i.e.\ from noise to the statistics of natural images. The correlation statistics extend to time.}
		\label{fig:colorbin1234stimstat}
	\end{center}
\end{figure}

 \begin{figure}[tbp]
	\begin{center}
		\includegraphics[width=0.32\textwidth]{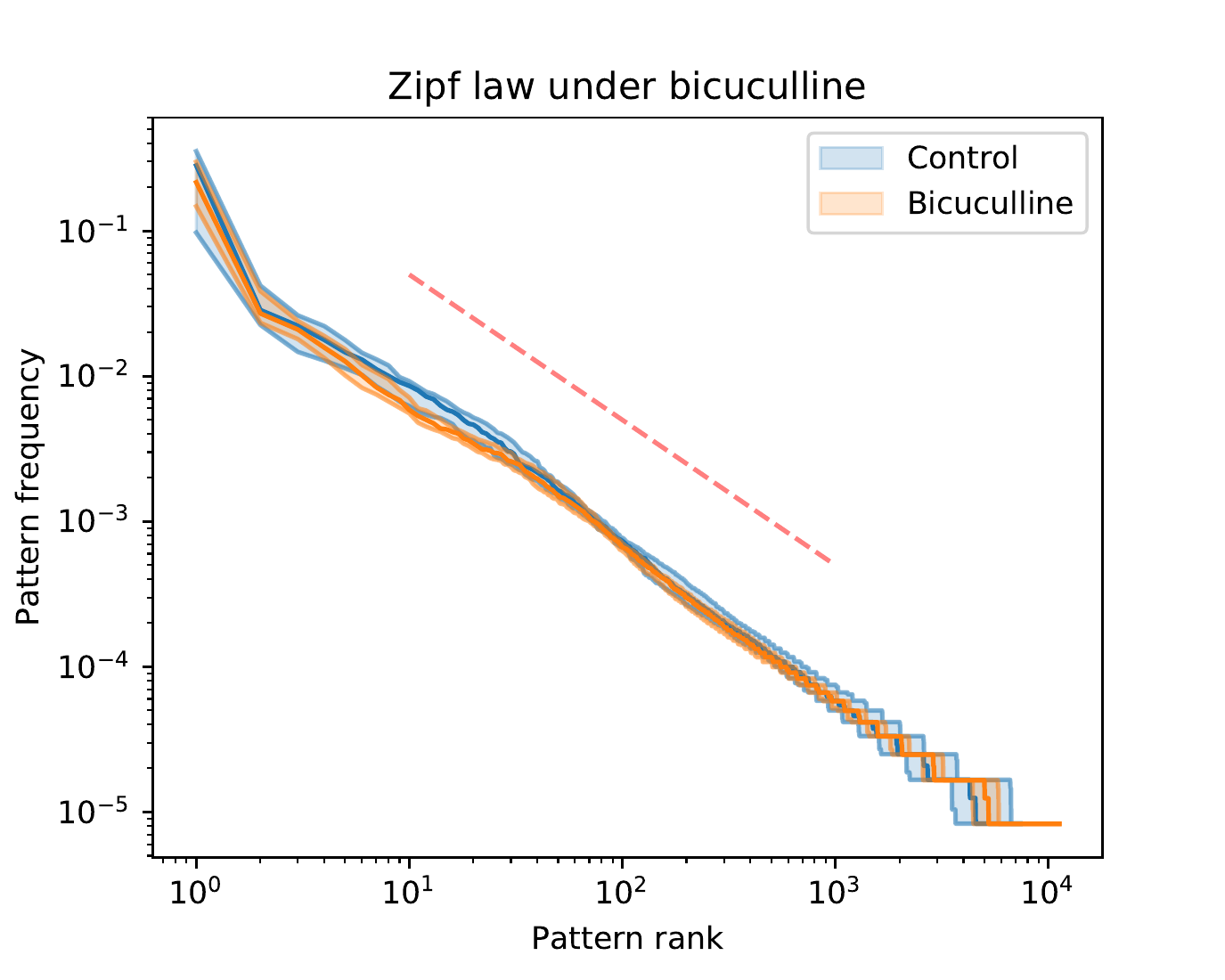}
		\includegraphics[width=0.32\textwidth]{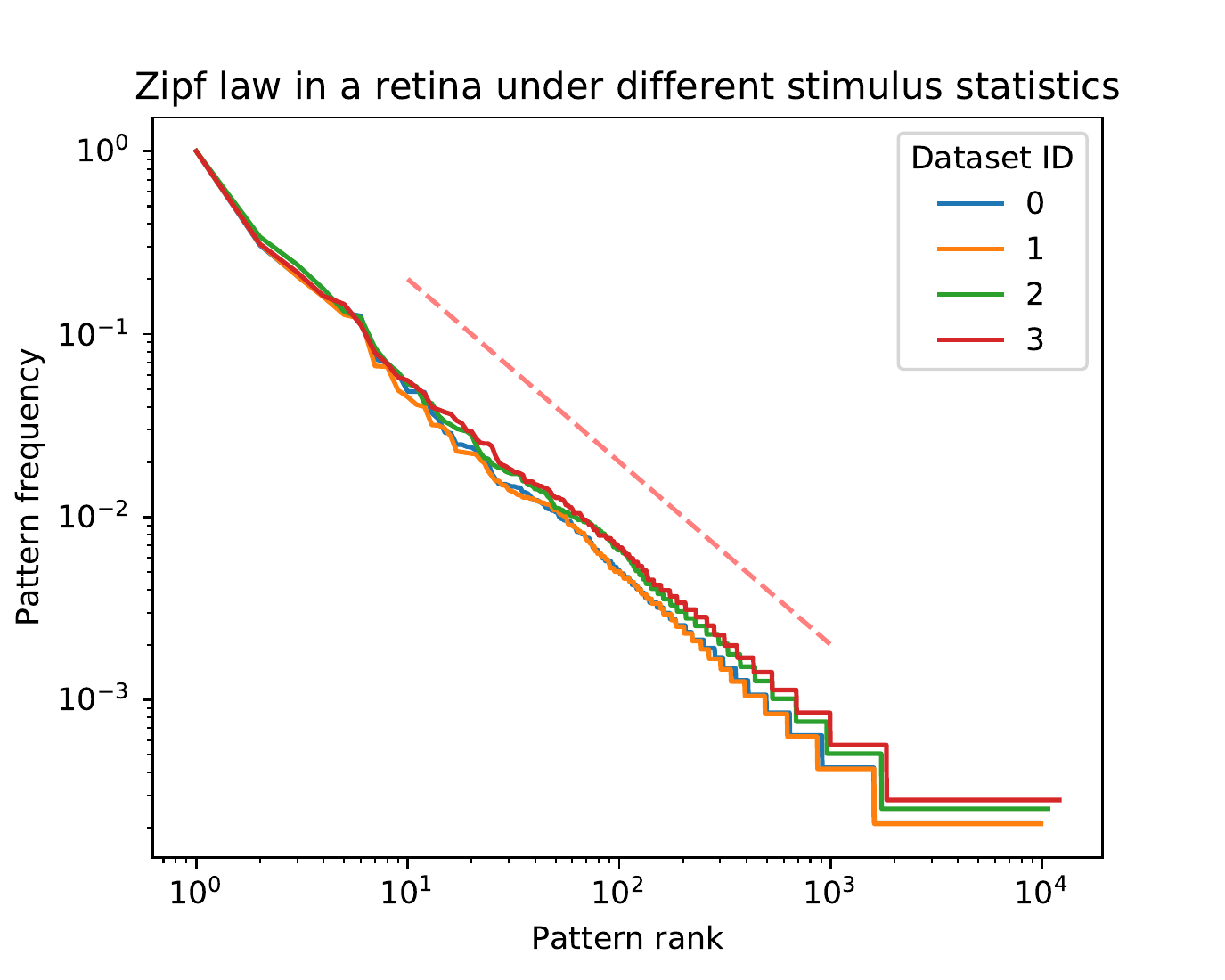}
		\includegraphics[width=0.32\textwidth]{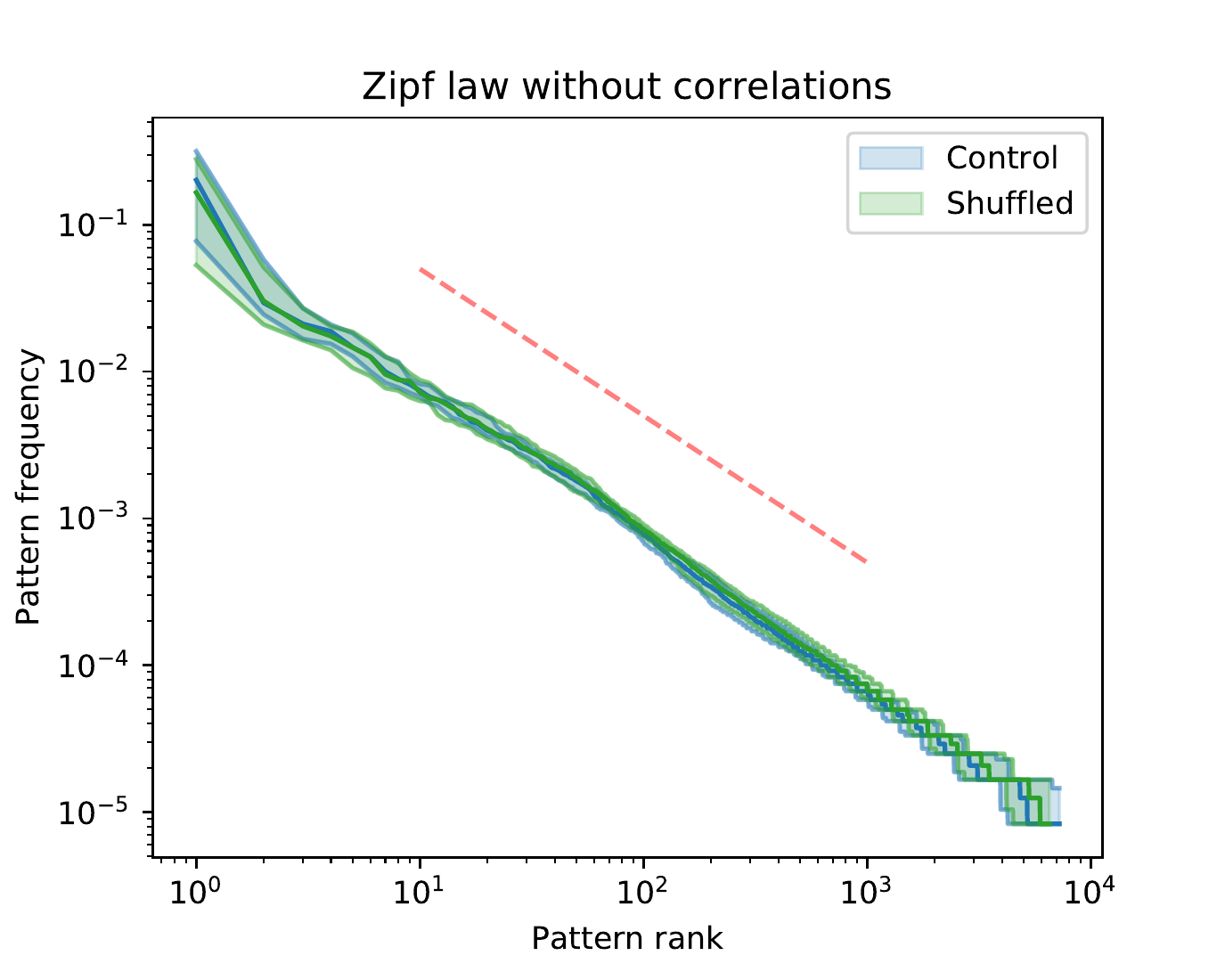}
		\caption{Left: Zipf plots, before and after treatment with bicuculline. 30 groups of 100 neurons, selected as explained in the Methods paragraph. Centre: Zipf plots for a unique group of 100 neurons under stimuli of different statistics; the difference between datasets 0-3 consist in the different spatial frequency --- from near-white noise to natural stimulus statistics. Right: the same data as in the left panel (control), and its shuffled version, where correlations have been destroyed, while keeping the same firing rates. The red dashed lines correspond to $1/x$ laws.}
		\label{fig:zipfs}
	\end{center}
\end{figure}

If Zipf's laws have a functional role, there is no expectation this phenomenon would survive in a non functioning neural system, such as a retina that has been pharmacologically treated in a way that breaks its normal operative mode. Here, we took data from the same mouse retina, before and after treatment with a 20 $\mu$M solution of bicuculline, which is a \textsc{gaba}\textsubscript{A} antagonist. The results are shown in figure \ref{fig:zipfs} (left): as it is evident, there is no clear difference between the two rank-probability plots. Of course, the only strong argument against the functional role of Zipf laws would be finding a functional retina in which this law is broken, which is not the case here. However, we can notice that even an intervention that significantly disrupts the retina's activity, by blocking inhibitory interactions, doesn't prevent this phenomenon to arise. This is despite the large change in the correlation between neurons induced by bicuculline.

Likewise, one may expect a dependence of pattern frequency-rank statistics on stimulus statistics. The retina, after all, is a neural system design to encode a stimulus --- and the correlation structure of its neurons' activity strongly depends on the correlations in the stimulus. However, we found no significant difference in Zipf laws under different stimulations. Figure \ref{fig:zipfs} (centre) shows a single group of 100 neurons selected in a retina that was stimulated with light patterns of different kinds. All stimulus presentations consisted of black-and-white random checkerboards, which are binarised versions of random noise of given frequency spectrum $f^{-a}$ with $a = 0.5, 1.0, 1.5, 2.0$ in space and time: from near-white noise to the statistics of natural images (figure \ref{fig:colorbin1234stimstat}).

The independence from correlations is evident in the right panel of figure \ref{fig:zipfs}: here, the ``control'' curve is the same as in the left panel, and is compared with the rank-probability plot for a ``shuffled'' version of the same data, where the firing rates were kept the same, but spikes were moved in time in order to cancel neuron-neuron correlations. The difference between the two curves is clearly not significant. This demonstrates how a firing rate distribution which is long-tailed (approximately log-normal) can in itself produce a Zipf-like plot. More research is needed to show whether this holds in general.

\paragraph*{Methods}
 
  \begin{figure}[tbp]
	\begin{center}
		\includegraphics[width=0.6\textwidth]{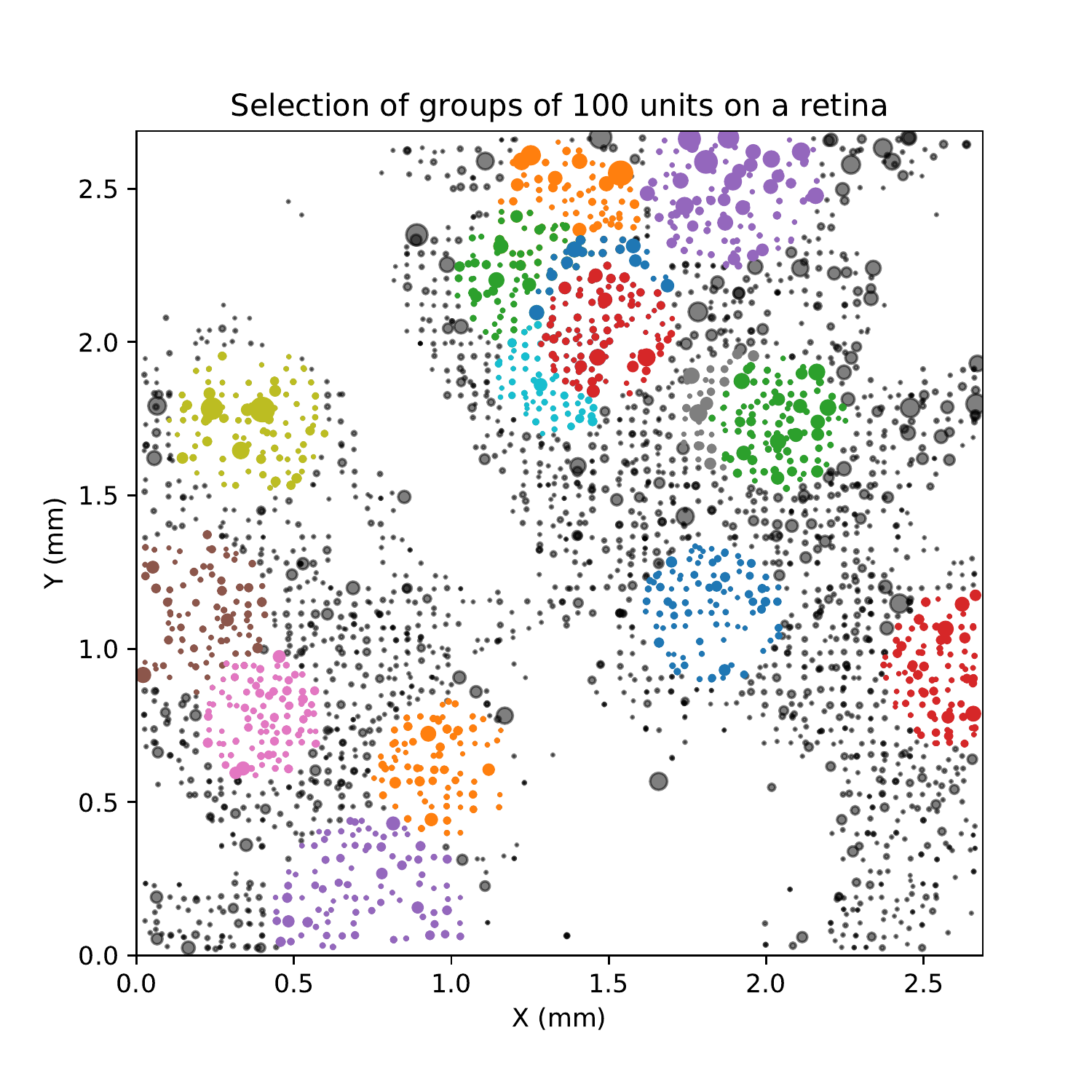}
		\caption{All spike clusters in a dataset (P91 mouse retina under white noise checkerboard stimulation), arranged spatially. For Zipf analysis, a random cluster was selected, and the 100 nearest ones picked along it (coloured patches on the figure are examples) to form a 100-neuron group. The process was repeated 30 times to study error intervals. The size of the dot scales with the number of spikes in the cluster. Even if this image only represents detected spikes, the optic disc is noticeable at the bottom end; other inactive areas corresponds to cuts in the retina, unavoidable when placing it on a flat surface.}
		\label{fig:retinal-patches}
	\end{center}
\end{figure}
 
The handling of the retinas, experimental apparatus, and the first part of the data analysis pipeline were performed as illustrated in \cite{hilgen2017unsupervised}. Starting from detected and sorted spikes, we removed those with very low amplitudes, by selecting a threshold corresponding roughly to the lowest 10\%. This was to ensure only good-quality events were left. Then, we selected, for each Zipf plot, $N=100$ clusters all pertaining to the same area of the retina (figure \ref{fig:retinal-patches}).

At this stage, spikes were binarised into a $N\times T$ matrix $S$ of boolean variables, with $S(n,t)=1$ if neuron $n$ spiked between times $t$ and $t+\delta t$ and $S(n, t) = 0$ otherwise. When multiple spikes from the same neuron occurred in a single time bin, the extra spikes were disregarded. For recordings shown in this chapter, $T=120 000$ or more, and $\delta t = 10$ ms, implying at least 20 minutes of neural activity were recorded.

\section{Parametric sensitivity\label{sec:fim}}

The basic fitting procedure of a maximum entropy model minimises
the quadratic difference between the data moments and the moments predicted 
by the model. During fitting, any model is updated by exploring the parameter space, following a direction given by the loss function. When a model admits a phase transition, the parameter space is characterised by (at least) two regions, corresponding to the phases, separated by a critical surface. From a theoretical point of view, asking why a model is poised at criticality coincides with asking why the fitting process tends to lead towards the critical surface in the parameter space. This has been discussed by \cite{mastromatteo2011criticality}; before introducing their argument, we provide some theoretical background.


\subsection{Model distance}
Intuitively, a phase transition occurs at a location (the critical point or critical surface) where an arbitrarily small change in the parameters yields a sharp, qualitative change in the behaviour of the model. In this section, we will formalise this idea, and link the notion of model distance to the statistical physics framework that we have introduced above.

A common measure of the distance (in model space) of a probability distribution $p$ from a given one $q$, both defined on a set $S$, is the Kullback-Leibler divergence
$$ D_{KL}(p; q) = \int_S p(x)\log\frac{p(x)}{q(x)}dx.$$
It measures the amount of information that is lost when approximating $p$ by $q$. The name \emph{divergence} stresses that this quantity does not have the mathematical properties of a distance, namely not being symmetric. If the model space is parametrised by ${\theta}$, and $p$ and $q$ are close to each other in this space, so that $q = P_\theta$ and $p = P_{\theta + \delta\theta}$, at second order in $\delta\theta$, the divergence can be approximated as
$$ D_{KL}(P_{\theta + \delta\theta}; P_\theta) \approx \frac{1}{2} \delta\theta^T F(\theta) \delta\theta, $$
where $F$ is called Fisher information tensor (FIT) which is given by
$$ F_{ij}(\theta) = -\int P_\theta(x) \frac{\partial \log P_\theta(x)}{\partial \theta_i} \frac{\partial  \log P_\theta(x)}{\partial \theta_j}  dx.$$
Fisher information is here expressed as a statistical quantity, but it has an important relation to the physics of statistical models. Consider a Hamiltonian model, where the probability distribution is given by
\begin{equation} P_\theta(x) = \frac{e^{-H_\theta(x)}}{Z_\theta}, \qquad H_\theta(x) = \sum_{k=1}^n\theta_kf_k(x), \label{this_form} \end{equation}
which is an obvious generalisation of maximum entropy models.
Calculating the Fisher information for this form of $P$ (\ref{this_form}),
we retrieve the direct connection between the covariance (with respect to 
$P_\theta$) of the physical quantities $f$ and the FIT that is characteristic
for probability distributions of the exponential type:
$$ F_{ij}(\theta) = \operatorname{Cov}[f_i(x), f_j(x)]. $$
This means that the FIT characterises the variances and correlations of the functions~$f$ which are now considered as stochastic variables and depend on 
the state $x$ of the system.

Additionally, note that changing the temperature in the traditional canonical ensemble corresponds to scaling the Hamiltonian by a factor $\beta = 1/T$, similarly to equation~(\ref{Ising_T}).
In the formulation above (\ref{this_form}), this is equivalent to scaling all the $\theta_i$ by $\beta$. Given a point $\theta$ on the parameter manifold, the direction $\partial/\partial\beta$ can be expressed as
$$ \frac{\partial}{\partial\beta} = \frac{1}{n}\sum_{k=1}^n \frac{\partial}{\partial\theta_k} $$
which is just a linear combination. The specific heat is given by
$$ c(\beta) = \frac{\beta^2}{N} \operatorname{Var}[E]. $$
Thus, we can arrange for the specific heat to be one of the entries of the Fisher information matrix, with a change of basis, which includes the $\beta$ direction as a base vector together with other $n-1$ linearly independent ones. Analogous considerations can be made for the magnetic field and magnetic susceptibility. In this sense, the Fisher information tensor is a generalisation of specific heats and susceptibilities. 

\subsection{Fisher information and criticality}
\label{sec:fishcrit}

We can now look at the relationship between statistical criticality and the model's parameter space. Suppose any \emph{generalised susceptibility} (i.e.~a component of the Fisher tensor) diverges at a point $\theta_0$. Then an eigenvalue of the Fisher information, say $\lambda_k$, diverges at $\theta_0$. Call $v_k$ the corresponding normalised eigenvector. For small $\alpha$,
$$ D_{KL}(P_{\theta_0+\alpha v_k}; P_{\theta_0}) \approx \frac{\alpha^2}{2}v_k^TF(\theta_0)v_k = \frac{\alpha^2}{2}\lambda_k, $$
and the r.h.s. diverges. This means that, moving from $\theta_0$ in the $v_k$ direction by an arbitrarily small step yields a model $P_{\theta_0+\alpha v_k}$ that is completely different from $P_{\theta_0}$, as indicated by an infinite KL divergence.

We introduced this description in terms of Fisher information in order to give an interpretation of criticality from the point of view of modelling. A model is at a critical point whenever there is a direction in parameter space that leads to an infinitely different model by a finite change in parameters. This, incidentally, shows that the best way of measuring the distance from a critical point is not to vary temperature, but to use the first eigenvalue of the FIT and move in the direction of the corresponding eigenvector. Temperature is not always the most relevant control parameter.

Mastromatteo and Marsili~\cite{mastromatteo2011criticality} have argued that, because of this special property critical points have in the parameter space, they are particularly favoured by model fitting. In particular, they show that \emph{distinguishable} models accumulate near critical points, whereas models farther away from these are largely indistinguishable. Their argument, in brief, goes a follows. Two models are considered indistinguishable if their Kullback-Leibler divergence is less than a given value $\epsilon$. For small $\epsilon$, $D_{KL}$ is approximated by Fisher information, and the volume of parameter space occupied by models indistinguishable from $\theta_0$ turns out to be proportional to $(\det F(\theta_0))^{-\frac{1}{2}}$. This quantity diverges at critical points due to the first eigenvalue diverging as explained above. Thus, \emph{most models} actually are poised near a critical point, according to this metric. They conclude that criticality may be a feature induced by the inference process, rather than one intrinsic to the real system being studied by the model. This may be the reason why statistical models seem to be poised at a critical point, for a variety of training datasets, as we showed in section \ref{sec:rbm}. However, it does not affect Zipf laws, which are directly observed in the data.


\subsection{Criticality and parameter `sloppiness'}

It is well known that the parameter spaces of many models often show only a small number of directions (linear combinations of parameter changes) along which the overall properties of the model strongly change (``stiff''), and a large number of directions which have little influence on the model (``sloppy''). This phenomenon, termed ``model sloppiness'', has been observed in a wide number of cases in systems science \cite{gutenkunst2007universally, machta2013parameter}.

For the specific case of neuronal networks, in Ref.~\cite{panas2015sloppiness}, although for small numbers of neurons, ``stiff'' dimensions corresponding to large FIT eigenvalues were identified. The remaining ``sloppy'' dimensions, on the other hand, can change without much effect on the goodness of fit of the model. A further development of this approach has been reported in Ref.~\cite{herzog2018dimensionality}, where it was shown that about half of the dimensions in the data manifold are irrelevant for the modelling. As shown in paragraph 4.2, near a critical point, the direction pointing towards the critical surface has a diverging FIT eigenvalue, while the others are smaller. This hints there may be a connection between sloppiness and criticality, which, at the moment, we can only leave at the level of speculation.



Additionally, however, sloppiness indicates that a fitting algorithm for the data may be improved if different dimensions are differently weighted during the optimisation process. We can then ask whether using a natural gradient in the fitting procedure would lead to a different result while evaluating model criticality. In natural gradient optimisation, the components of the gradient are compensated by the inverse Fisher information, i.e.~the divergence near a critical point of the model would disappear, at least theoretically when the Fisher information is exactly known. As a result, the fitting procedure is not homogeneous with respect to the set of the parameters, but with respect to the space of the parameters, taking into account its geometrical properties, and parameters can be identified equally well in all regions. In this way, the problem discussed in Ref.~\cite{mastromatteo2011criticality} may disappear --- more research will be needed in order to verify this.


\section{Discussion\label{Discussion}}

Neuronal avalanches are an experimentally well-studied phenomenon, that
can be explained as a consequence of the optimisation of information
processing in the brain. It should be noted that 
an understanding of how the potential functional
benefits of this ``dynamical'' criticality are realised is missing~\cite{shew2013functional} 
 --- however, it has been shown that the maximisation 
of the dynamical range happens at criticality~\cite{Kinouchi2006}.

Statistical criticality is an equally complex phenomenon to explain theoretically. Like dynamical criticality, it can be taken to indicate the complexity of the neural data and the relevance of higher-order correlations or latent variables, but its functional implications are less clear. In this chapter, we have reviewed the concept, both in the context of fitted statistical models, and as a direct observation of Zipf laws in neural population data. Through experiments on restricted Boltzmann machines, we suggested that the divergence of model specific heat is not a reliable way to infer properties of the data.  We mentioned how Fisher information provides the correct description of the parameter space and the critical surfaces, and reviewed a possible explanation of why statistical models tend to poise themselves at a critical point.  Then, we tried to describe the connection between statistical and dynamical criticality, and argued there is no clear connection, by showing examples where one of the two was present without the other. Further insight on this matter might come from models that are capable of both, provided they can reproduce not only the equilibrium distribution of the data, but also the dynamics. A multi-time maximum entropy model might provide a starting point for this work.

Of course, it may well be that the observation of Zipf laws is simply a consequence of problems related to how we describe the data --- these include the typically small sets of observables, the choice of binning size, failure to account for the real dynamics, and biases introduced by sampling. However, the ubiquity of Zipf laws in complex systems means its emergence in biological neural networks should not surprise us, and it could be explained in terms of mechanisms such as the one described by \cite{aitchison2016zipf}, or perhaps with preferential attachment. Conversely, an important open problem is an explanation on whether statistical criticality is something that is actively sought by the system because of some functional relevance. On this matter, we tried to analyse the Zipf profile of retinal activity under various conditions (various stimulus statistics, pharmacological treatment), but we found no significant differences in the cases examined. Interestingly, it seems to be possible to generate a Zipf profile simply by enforcing a long-tailed firing rate distribution, despite the absence of correlations. Even if this observation were confirmed, the question would simply shift towards finding a reason for such a skewed distribution of firing rates, which has not yet found a justification in terms of function.

Notably, recent research has started showing how Zipf laws appear in different kinds of parametric models, including ``deep'' ones, as soon as learning occurs. It has been shown that the Zipf property arises to different degrees in different layers of a deep network, and is maximal in the layers that attain an optimal trade-off between resolution and accuracy in generating samples \cite{song2017emergence}. This is a starting point in linking statistical criticality to function. It is not known whether similar principles are relevant in the case of biological neural networks, and finding such a link could be an interesting direction of future research.

\bibliographystyle{spmpsci}
\bibliography{thebib3,thebib3-michael}

\begin{thebibliography}{10}
\providecommand{\url}[1]{{#1}}
\providecommand{\urlprefix}{URL }
\expandafter\ifx\csname urlstyle\endcsname\relax
  \providecommand{\doi}[1]{DOI~\discretionary{}{}{}#1}\else
  \providecommand{\doi}{DOI~\discretionary{}{}{}\begingroup
  \urlstyle{rm}\Url}\fi

\bibitem{aitchison2016zipf}
Aitchison, L., Corradi, N., Latham, P.E.: Zipf’s law arises naturally when
  there are underlying, unobserved variables.
\newblock PLoS Computational Biology \textbf{12}(12), 1--32 (2016)

\bibitem{Athreyab}
Athreya, K.B., Jagers, P.: Classical and Modern Branching Processes,
  \emph{IMA}, vol.~84.
\newblock Springer (1997)

\bibitem{Auerbach1913}
Auerbach, F.: Das {G}esetz der {B}ev{\"o}lkerungskonzentration.
\newblock Petermanns Geographische Mitteilungen \textbf{59}, 74--76 (1913).
\newblock (Quote translated by J.M.H.)

\bibitem{Barabasi1999}
Barab\'asi, A.L., Albert, R.: Emergence of scaling in random networks.
\newblock Science \textbf{286}, 509--512 (1999)

\bibitem{beggs2008criticality}
Beggs, J.M.: The criticality hypothesis: how local cortical networks might
  optimize information processing.
\newblock Philosophical Transactions of the Royal Society of London A:
  Mathematical, Physical and Engineering Sciences \textbf{366}(1864), 329--343
  (2008)

\bibitem{beggs2003neuronal}
Beggs, J.M., Plenz, D.: Neuronal avalanches in neocortical circuits.
\newblock Journal of Neuroscience \textbf{23}(35), 11,167--11,177 (2003)

\bibitem{beggs2012being}
Beggs, J.M., Timme, N.: Being critical of criticality in the brain.
\newblock Frontiers in Physiology \textbf{3}, 163 (2012)

\bibitem{Cristelli2012}
Cristelli, M., Batty, M., Pietronero, L.: There is more than power law in
  {Z}ipf.
\newblock Scientific Reports \textbf{2}, 812(7) (2012)

\bibitem{eurich2002finite}
Eurich, C.W., Herrmann, J.M., Ernst, U.A.: Finite-size effects of avalanche
  dynamics.
\newblock Physical Review E \textbf{66}(6), 066,137 (2002)

\bibitem{gabaix1999zipf}
Gabaix, X.: Zipf's law and the growth of cities.
\newblock American Economic Review \textbf{89}(2), 129--132 (1999)

\bibitem{gardella2018blindfold}
Gardella, C., Marre, O., Mora, T.: Blindfold learning of an accurate neural
  metric.
\newblock Proceedings of the National Academy of Sciences p. 201718710 (2018)

\bibitem{gautam2015maximizing}
Gautam, S.H., Hoang, T.T., McClanahan, K., Grady, S.K., Shew, W.L.: Maximizing
  sensory dynamic range by tuning the cortical state to criticality.
\newblock PLoS Computational Biology \textbf{11}(12), e1004,576 (2015)

\bibitem{Glauber1963}
Glauber, R.J.: Time-dependent statistics of the {I}sing model.
\newblock Journal of Mathematical Physics \textbf{4}(2), 294--307 (1963)

\bibitem{gutenkunst2007universally}
Gutenkunst, R.N., Waterfall, J.J., Casey, F.P., Brown, K.S., Myers, C.R.,
  Sethna, J.P.: Universally sloppy parameter sensitivities in systems biology
  models.
\newblock PLoS Computational Biology \textbf{3}(10), e189 (2007)

\bibitem{hahn2017spontaneous}
Hahn, G., Ponce-Alvarez, A., Monier, C., Benvenuti, G., Kumar, A., Chavane, F.,
  Deco, G., Fr{\'e}gnac, Y.: Spontaneous cortical activity is transiently
  poised close to criticality.
\newblock PLoS Computational Biology \textbf{13}(5), 1--29 (2017)

\bibitem{hennig2009early}
Hennig, M.H., Adams, C., Willshaw, D., Sernagor, E.: Early-stage waves in the
  retinal network emerge close to a critical state transition between local and
  global functional connectivity.
\newblock The Journal of Neuroscience \textbf{29}(4), 1077--1086 (2009)

\bibitem{herzog2018dimensionality}
Herzog, R., Escobar, M.J., Cofre, R., Palacios, A.G., Cessac, B.:
  Dimensionality reduction on spatio-temporal maximum entropy models on spiking
  networks.
\newblock Preprint \emph{bioRxiv}:278606  (2018)

\bibitem{hilgen2017unsupervised}
Hilgen, G., Sorbaro, M., Pirmoradian, S., Muthmann, J.O., Kepiro, I.E., Ullo,
  S., Ramirez, C.J., Encinas, A.P., Maccione, A., Berdondini, L., et~al.:
  Unsupervised spike sorting for large-scale, high-density multielectrode
  arrays.
\newblock Cell reports \textbf{18}(10), 2521--2532 (2017)

\bibitem{Hopfield1982}
Hopfield, J.J.: Neural networks and physical systems with emergent collective
  computational abilities.
\newblock Proceedings of the National Academy of Sciences \textbf{79}(8),
  2554--2558 (1982)

\bibitem{ising1925beitrag}
Ising, E.: Beitrag zur {T}heorie des {F}erromagnetismus.
\newblock Zeitschrift f{\"u}r Physik \textbf{31}(1), 253--258 (1925)

\bibitem{jaynes1957information}
Jaynes, E.T.: Information theory and statistical mechanics.
\newblock Physical Review \textbf{106}(4), 620--630 (1957)

\bibitem{jiang2011zipf}
Jiang, B., Jia, T.: Zipf's law for all the natural cities in the united states:
  a geospatial perspective.
\newblock International Journal of Geographical Information Science
  \textbf{25}(8), 1269--1281 (2011)

\bibitem{Kinouchi2006}
Kinouchi, O., Copelli., M.: Optimal dynamical range of excitable networks at
  criticality.
\newblock Nat. Phys. \textbf{2}, 348--352 (2006)

\bibitem{koster2014modeling}
K{\"o}ster, U., Sohl-Dickstein, J., Gray, C.M., Olshausen, B.A.: Modeling
  higher-order correlations within cortical microcolumns.
\newblock PLoS Computational Biology \textbf{10}(7), e1003,684 (2014)

\bibitem{larremore2011predicting}
Larremore, D.B., Shew, W.L., Restrepo, J.G.: Predicting criticality and dynamic
  range in complex networks: effects of topology.
\newblock Physical Review Letters \textbf{106}(5), 058,101 (2011)

\bibitem{Li1992}
Li, W.: Random texts exhibit {Z}ipf's-law-like word frequency distribution.
\newblock IEEE Transactions on Information Theory \textbf{38}(6), 1842--1845
  (1992)

\bibitem{machta2013parameter}
Machta, B.B., Chachra, R., Transtrum, M.K., Sethna, J.P.: Parameter space
  compression underlies emergent theories and predictive models.
\newblock Science \textbf{342}(6158), 604--607 (2013)

\bibitem{marre2009prediction}
Marre, O., El~Boustani, S., Fr{\'e}gnac, Y., Destexhe, A.: Prediction of
  spatiotemporal patterns of neural activity from pairwise correlations.
\newblock Physical Review Letters \textbf{102}(13), 138,101 (2009)

\bibitem{mastromatteo2011criticality}
Mastromatteo, I., Marsili, M.: On the criticality of inferred models.
\newblock Journal of Statistical Mechanics: Theory and Experiment
  \textbf{2011}(10), P10,012 (2011)

\bibitem{mizuseki2013preconfigured}
Mizuseki, K., Buzs{\'a}ki, G.: Preconfigured, skewed distribution of firing
  rates in the hippocampus and entorhinal cortex.
\newblock Cell reports \textbf{4}(5), 1010--1021 (2013)

\bibitem{mora2015dynamical}
Mora, T., Deny, S., Marre, O.: Dynamical criticality in the collective activity
  of a population of retinal neurons.
\newblock Physical Review Letters \textbf{114}(7), 078,105 (2015)

\bibitem{nasser2013spatio}
Nasser, H., Marre, O., Cessac, B.: Spatio-temporal spike train analysis for
  large scale networks using the maximum entropy principle and monte carlo
  method.
\newblock Journal of Statistical Mechanics: Theory and Experiment
  \textbf{2013}(03), P03,006 (2013)

\bibitem{newman2005power}
Newman, M.E.: Power laws, {P}areto distributions and {Z}ipf's law.
\newblock Contemporary physics \textbf{46}(5), 323--351 (2005)

\bibitem{nishimori2001statistical}
Nishimori, H.: Statistical physics of spin glasses and information processing:
  an introduction, vol. 111.
\newblock Clarendon Press (2001)

\bibitem{nonnenmacher2017signatures}
Nonnenmacher, M., Behrens, C., Berens, P., Bethge, M., Macke, J.H.: Signatures
  of criticality arise from random subsampling in simple population models.
\newblock PLoS Computational Biology \textbf{13}(10), e1005,718 (2017)

\bibitem{odonnell2016population}
O'Donnell, C., Gon{\c{c}}alves, J.T., Whiteley, N., Portera-Cailliau, C.,
  Sejnowski, T.J.: The population tracking model: A simple, scalable
  statistical model for neural population data.
\newblock Neural Computation \textbf{29}(1), 50--93 (2016)

\bibitem{ohiorhenuan2010sparse}
Ohiorhenuan, I.E., Mechler, F., Purpura, K.P., Schmid, A.M., Hu, Q., Victor,
  J.D.: Sparse coding and high-order correlations in fine-scale cortical
  networks.
\newblock Nature \textbf{466}(7306), 617--621 (2010)

\bibitem{panas2015sloppiness}
Panas, D., Amin, H., Maccione, A., Muthmann, O., van Rossum, M., Berdondini,
  L., Hennig, M.H.: Sloppiness in spontaneously active neuronal networks.
\newblock Journal of Neuroscience \textbf{35}(22), 8480--8492 (2015)

\bibitem{priesemann2013neuronal}
Priesemann, V., Valderrama, M., Wibral, M., Le~Van~Quyen, M.: Neuronal
  avalanches differ from wakefulness to deep sleep--evidence from intracranial
  depth recordings in humans.
\newblock PLoS Computational Biology \textbf{9}(3), e1002,985 (2013)

\bibitem{redner1998popular}
Redner, S.: How popular is your paper? an empirical study of the citation
  distribution.
\newblock The European Physical Journal B-Condensed Matter and Complex Systems
  \textbf{4}(2), 131--134 (1998)

\bibitem{schneidman2006weak}
Schneidman, E., Berry, M.J., Segev, R., Bialek, W.: Weak pairwise correlations
  imply strongly correlated network states in a neural population.
\newblock Nature \textbf{440}(7087), 1007--1012 (2006)

\bibitem{shew2013functional}
Shew, W.L., Plenz, D.: The functional benefits of criticality in the cortex.
\newblock The Neuroscientist \textbf{19}(1), 88--100 (2013)

\bibitem{shew2009neuronal}
Shew, W.L., Yang, H., Petermann, T., Roy, R., Plenz, D.: Neuronal avalanches
  imply maximum dynamic range in cortical networks at criticality.
\newblock Journal of Neuroscience \textbf{29}(49), 15,595--15,600 (2009)

\bibitem{shlens2006structure}
Shlens, J., Field, G.D., Gauthier, J.L., Grivich, M.I., Petrusca, D., Sher, A.,
  Litke, A.M., Chichilnisky, E.: The structure of multi-neuron firing patterns
  in primate retina.
\newblock The Journal of Neuroscience \textbf{26}(32), 8254--8266 (2006)

\bibitem{song2017emergence}
Song, J., Marsili, M., Jo, J.: Emergence and relevance of criticality in deep
  learning.
\newblock arXiv preprint arXiv:1710.11324  (2017)

\bibitem{Tang2008}
Tang, A., Jackson, D., Hobbs, J., Chen, W., Smith, J.L., Patel, H., Prieto, A.,
  Petrusca, D., Grivich, M.I., Sher, A., Hottowy, P., Dabrowski, W., Litke,
  A.M., Beggs, J.M.: A maximum entropy model applied to spatial and temporal
  correlations from cortical networks in vitro.
\newblock Journal of Neuroscience \textbf{28}, 505–518 (2008)

\bibitem{tkacik2014searching}
Tka{\v{c}}ik, G., Marre, O., Amodei, D., Schneidman, E., Bialek, W., Berry~II,
  M.J.: Searching for collective behavior in a large network of sensory
  neurons.
\newblock PLoS Computational Biology \textbf{10}(1), e1003,408 (2014)

\bibitem{tkacik2015thermodynamics}
Tka{\v{c}}ik, G., Mora, T., Marre, O., Amodei, D., Palmer, S.E., Berry, M.J.,
  Bialek, W.: Thermodynamics and signatures of criticality in a network of
  neurons.
\newblock Proceedings of the National Academy of Sciences \textbf{112}(37),
  11,508--11,513 (2015)

\bibitem{vazquez2017stochastic}
V{\'a}zquez-Rodr{\'\i}guez, B., Avena-Koenigsberger, A., Sporns, O., Griffa,
  A., Hagmann, P., Larralde, H.: Stochastic resonance at criticality in a
  network model of the human cortex.
\newblock Scientific Reports \textbf{7}(1), 13,020 (2017)

\bibitem{Vitanov2015}
Vitanov, N.K., Ausloos, M.: Test of two hypotheses explaining the size of
  populations in a system of cities.
\newblock Journal of Applied Statistics \textbf{42}(12), 2686--2693 (2015)

\bibitem{Yu2011}
Yu, S., Yang, H., Nakahara, H., Santos, G.S., Nikoli\'c, D., Plenz, D.:
  Higher-order interactions characterized in cortical activity.
\newblock Journal of Neuroscience \textbf{30}(48), 17,514--17,526 (2011)

\bibitem{Zipf1949}
Zipf, G.K.: Human behavior and the principle of least effort.
\newblock Addison-Wesley, Cambridge (1949)

\end{thebibliography}
\end{document}